\documentclass[epj]{webofc}
\usepackage[utf8]{inputenc}
\usepackage[varg]{txfonts}   
\usepackage{booktabs}
\usepackage{xcolor}
\definecolor{darkred}{rgb}{0.4,0.0,0.0}
\definecolor{darkgreen}{rgb}{0.0,0.4,0.0}
\definecolor{darkblue}{rgb}{0.0,0.0,0.4}
\usepackage[bookmarks,linktocpage,colorlinks,
    linkcolor = darkred,
    urlcolor  = darkblue,
    citecolor = darkgreen]{hyperref}
\usepackage{subfigure}
\wocname{EPJ Web of Conferences}
\woctitle{Lattice2017}
\begin{document}
\begin{flushright} 
KEK-TH-2010
\end{flushright} 

\vspace{-0.3cm}

\selectlanguage{english}
\title{%
Comparative studies of the deformation techniques for the singular-drift 
problem in the complex Langevin method
}
\author{%
\firstname{Yuta} \lastname{Ito}\inst{1}\fnsep\thanks{Speaker, \email{yito@post.kek.jp}} \and
\firstname{Jun} \lastname{Nishimura}\inst{1,2}
}
\institute{%
KEK Theory Center, High Energy Accelerator Research Organization,\\
1-1 Oho, Tsukuba, Ibaraki 305-0801, Japan
\and
Graduate University for Advanced Studies (SOKENDAI),\\
1-1 Oho, Tsukuba, Ibaraki 305-0801, Japan
}
\abstract{%
  In application of the complex Langevin method to QCD at high density and 
  low temperature, the singular-drift problem occurs due to the appearance 
  of near-zero eigenvalues of the Dirac operator. In order to avoid this 
  problem, we proposed to deform the Dirac operator in such a way that the 
  near-zero eigenvalues do not appear and to extrapolate the deformation 
  parameter to zero from the available data points. Here we test three 
  different types of deformation in a simple large-$N$ matrix model, which 
  undergoes an SSB due to the phase of the fermion determinant, and compare 
  them to see the consistency with one another.
}
\maketitle

\section{Introduction\label{sec:Introduction}}
When the action is complex, the integrand
in the path integral cannot be regarded as the probability
distribution. Therefore, the usual Monte Carlo simulations cannot be
applied to such systems. A brute-force method is 
the so-called re-weighting method, but since the complex phase oscillates 
violently as the system size increases,
it is difficult to evaluate the expectation values using
this kind of method.

The complex Langevin method (CLM) \cite{Parisi:1984cs,Klauder:1983sp}
is a promising approach to this sign problem.
It can be viewed as a complex extension of the stochastic quantization
based on the Langevin equation. The idea of the CLM is to complexify
the dynamical variables and to consider the holomorphic extension of the
action and the observables. While the CLM works even
in some systems that suffer from a severe sign problem, it
gives simply wrong results in the other cases. 
One of the recent developments concerns
the conditions required for justification of the method.
It was found that the probability distribution of the drift term
has to fall off faster than exponential \cite{Nagata:2016vkn}.
There are two cases in which this is not satisfied.
One is the case in which 
the dynamical variables make long excursions 
into the imaginary direction \cite{Aarts:2009uq}.
The other is the case in which
there is a non-vanishing probability that dynamical variables
come close to the singularity of the drift term 
(the singular-drift problem) \cite{Nishimura:2015pba}.
Another development was the invention of 
the gauge cooling technique \cite{Seiler:2012wz},
which made the CLM work in finite density QCD 
either at high temperature \cite{Aarts:2014bwa}
or in the heavy dense limit \cite{Sexty:2013ica}.

At low temperature and high density, on the other hand,
it is anticipated that the singular-drift problem occurs
due to the appearance of near-zero eigenvalues of the Dirac operator. 
In ref.~\cite{Ito:2016efb},
we proposed to avoid this problem by deforming
the Dirac operator and extrapolating the deformation parameter to zero
using only the reliable results obtained by the deformed model.
We tested this idea in an SO(4)-symmetric matrix model with a Gaussian
action and a complex fermion determinant, in which spontaneous breaking
of the SO(4) symmetry is expected to occur due 
to the phase of the determinant \cite{Nishimura:2001sq}.
This is also confirmed by explicit calculations
based on the Gaussian expansion method (GEM) \cite{Nishimura:2004ts}.
Applying the CLM to this model,
we have found that the singular-drift problem is actually severe because
the fermionic part of the model is essentially an exactly ``massless''
system. Following the idea described above, we have found that the
SO(4) symmetry of the original matrix model is broken spontaneously
to SO(2) after extrapolating the deformation parameter to zero.
The obtained results were indeed consistent with the prediction of the GEM
albeit with small discrepancies.

There are, however, a few issues that remain to be addressed
in this deformation technique.
First the validity of the extrapolation relies on the assumption
that there is no phase transition in the parameter region in which the
CLM does not work. Second it is not straightforward to 
estimate the systematic errors associated with the extrapolation.
In order to address these issues, it is useful to 
think of various ways to deform the Dirac operator
and to see how much the extrapolated results depend on the types of deformation.
Here we investigate the aforementioned
SO(4)-symmetric matrix model using
three different types of deformed Dirac operator.
We find that the results obtained with these deformations
are consistent with one another, which supports the validity
and usefulness of the deformation technique as a solution to the
singular-drift problem in the CLM.

The rest of this paper is organized as follows. 
In section \ref{sec:Brief-review-of},
we review the SO(4)-symmetric matrix model that we investigate in this work.
In section \ref{sec:Complex-Langevin-simulations},
we explain how we apply the CLM to the SO(4)-symmetric matrix model.
In particular, we define three different types of deformation, each
of which can avoid the appearance of near-zero eigenvalues of the
Dirac operator but with different ideas. In section \ref{sec:Results},
we present the results obtained with the three deformations
and show that the results after extrapolating the deformation parameter are 
consistent with one another.
We also conclude that the small discrepancies with the prediction from the GEM
are due to the approximation used in the GEM.
Section \ref{sec:Discussion} is devoted to a summary and discussions.

\section{Brief review of the SO(4)-symmetric matrix model \label{sec:Brief-review-of}}

The SO(4)-symmetric matrix model investigated in this paper is defined
by the partition function \cite{Nishimura:2001sq} 
\begin{equation}
Z=\int dX\,\left(\det D\right)^{N_{\mathrm{f}}}
e^{-S_{{\rm b}}} \ ,
\label{part-fn-with-det}
\end{equation}
where the bosonic part of the action is given as 
\begin{equation}
S_{{\rm b}} = 
\frac{1}{2}N\sum_{\mu=1}^{4}{\rm tr}\,(X_{\mu})^{2}\ .
\label{eq:boson_action}
\end{equation}
We have introduced $N\times N$ Hermitian matrices $X_{\mu}$
$(\mu=1,\ldots,4)$. 
%
The Dirac operator $D$ in eq.~(\ref{part-fn-with-det})
is defined by
\begin{equation}
D_{i\alpha,j\beta}=\sum_{\mu=1}^{4}(\Gamma_{\mu})_{\alpha\beta}(X_{\mu})_{ij} \ .
\label{eq:dirac_op}
\end{equation}
Here the $2\times2$ matrices $\Gamma_{\mu}$
are the gamma matrices in 4d Euclidean space after Weyl projection defined by 
\begin{equation}
\Gamma_{\mu}=\begin{cases}
i\,\sigma_{i} & \text{for}\;\mu=i=1,2,3\ ,\\
\mathbf{1}_{2} & \text{for~}\mu=4\ ,
\end{cases}
\end{equation}
where $\sigma_{i}$ ($i=1,2,3$) are the Pauli matrices.
The model has an SO(4) symmetry, under which $X_{\mu}$ transforms
as a vector. 

The fermion determinant $\det D$ in (\ref{part-fn-with-det})
is complex in general.
It was speculated that the SO(4) rotational symmetry of the model
is spontaneously broken in the large-$N$ limit with fixed $r=N_{{\rm f}}/N>0$
due to the effect of the phase of the fermion determinant \cite{Nishimura:2001sq}.
In the phase-quenched model, which is defined by omitting the phase
of the fermion determinant, the SSB was shown not to occur by Monte
Carlo simulation \cite{Anagnostopoulos:2011cn}.
We may therefore say that the SSB, if it really occurs, should be
induced by the phase of the fermion determinant. Throughout this paper,
we consider the $r=1$ case, which corresponds to $N_{{\rm f}}=N$.

In order to see the SSB, we introduce an SO(4)-breaking mass term
\begin{equation}
\Delta S_{{\rm b}}=\frac{N}{2}\varepsilon
\sum_{\mu=1}^{4}m_{\mu}{\rm tr}\,(X_{\mu})^{2}
\label{eq:boson_action_bdeform}
\end{equation}
in the action, where 
\begin{equation}
m_{1}<m_{2}<m_{3}<m_{4}\ ,
\label{eq:boson_mass_split}
\end{equation}
and we define the order parameters for the SSB by the expectation values of 
\begin{alignat}{1}
\lambda_{\mu}=\frac{1}{N}{\rm tr}\,(X_{\mu})^{2}\ ,
\label{lambda-def}
\end{alignat}
where no sum over $\mu$ is taken. 
Due to the ordering (\ref{eq:boson_mass_split}), the expectation values obey 
\begin{alignat}{1}
\langle\lambda_{1}\rangle>
\langle\lambda_{2}\rangle>\langle\lambda_{3}\rangle
>\langle\lambda_{4}\rangle
\label{lambda-ordering}
\end{alignat}
at finite $\varepsilon$. Taking the large-$N$ limit and then sending
$\varepsilon$ to zero afterwards, the expectation values $\langle\lambda_{\mu}\rangle$
($\mu=1,\cdots,4$) may not take the same value. In that case, we
can conclude that the SSB occurs. Here and henceforth, the parameters
$m_{\mu}$ in the SO(4)-breaking term (\ref{eq:boson_action_bdeform})
are chosen as 
\begin{equation}
(m_{1},m_{2},m_{3},m_{4})=(1,2,4,8)\ .\label{eq:boson_mass_used}
\end{equation}

Explicit calculations based on the GEM were carried out assuming that
the SO(4) symmetry is broken down 
either to SO(2) or to SO(3) \cite{Nishimura:2004ts}.
For $r=1$, the order parameters are given by 
\begin{alignat}{1}
\left\langle \lambda_{1}\right\rangle 
=\left\langle \lambda_{2}\right\rangle 
\sim2.1\ ,\quad\left\langle \lambda_{3}\right\rangle 
\sim1.0\ ,\quad\left\langle \lambda_{4}\right\rangle 
\sim0.8\quad\quad & 
\mbox{for the \ensuremath{{\rm SO(2)}} vacuum}\ ,
\label{eq:previous_result}\\
\left\langle \lambda_{1}\right\rangle =
\left\langle \lambda_{2}\right\rangle =
\left\langle \lambda_{3}\right\rangle 
\sim1.75\ ,\quad\left\langle \lambda_{4}\right\rangle 
\sim0.75\quad\quad & 
\mbox{for the \ensuremath{{\rm SO(3)}} vacuum}\ .
\label{eq:previous_result_SO3}
\end{alignat}
The free energy was calculated in each vacuum, and the SO(2)-symmetric
vacuum was found to have the lower value. 

\section{Complex Langevin simulation with the deformed Dirac operator
\label{sec:Complex-Langevin-simulations}}

In this section, we explain how we apply the CLM to the SO(4)-symmetric
matrix model (\ref{part-fn-with-det}). Including the symmetry breaking
term (\ref{eq:boson_action_bdeform}), we can write the partition
function as 
\begin{alignat}{1}
Z & =\int dX\,w(X)\ ,\quad\quad 
w(X)=\left(\det D\right)^{N_{{\rm f}}}e^{-(S_{{\rm b}}+\Delta S_{{\rm b}})}\ .\label{part-fn-with-det-rewrite}
\end{alignat}
The drift term that appears in the Langevin equation 
is given by 
\begin{alignat}{1}
(v_{\mu})_{ij} & =
\frac{1}{w(X)}
\frac{\partial w(X)}{\partial(X_{\mu})_{ji}}
=-N\,\left(1+\varepsilon m_{\mu}\right)(X_{\mu})_{ij}
+N_{{\rm f}}\,(D^{-1})_{i\alpha,j\beta}(\Gamma_{\mu})_{\beta\alpha}
\label{drift-term}
\end{alignat}
as a function of the Hermitian matrices $X_{\mu}$. Note that the
second term in (\ref{drift-term}) is not Hermitian in general corresponding
to the fact that the fermion determinant is complex. Accordingly,
$X_{\mu}$ has to be extended to general complex matrices
as we solve the fictitious time evolution based on the Langevin equation
\footnote{In order to keep the matrices $X_{\mu}$ as close to Hermitian
as possible, we use the gauge cooling technique. 
Namely we define the Hermiticity norm 
${\cal N}_{{\rm H}}=\frac{1}{4N}\sum_{\mu=1}^{4}{\rm tr}
\left[\left(X_{\mu}-X_{\mu}^{\dagger}\right)
\left(X_{\mu}-X_{\mu}^{\dagger}\right)^{\dagger}\right]\ ,$
which measures the deviation of $X_{\mu}$ from a Hermitian configuration,
and minimize this norm using ${\rm SL}(N,\mathbb{C})$ transformation
after each Langevin step.}. 
The drift term (\ref{drift-term}) in the Langevin equation has to be
defined for general complex matrices $X_\mu$ by analytic continuation.

As we mentioned earlier, the singular-drift problem is associated
with the appearance of near-zero eigenvalues of the Dirac operator $D$.
Indeed we find that there are many eigenvalues close to zero
for $\varepsilon\leq 0.5$, which implies that the extrapolation to
$\varepsilon=0$ cannot be made reliably as it stands.
In order to circumvent this problem, we proposed to
deform the Dirac operator \cite{Ito:2016efb}.
Using this deformation technique, we have successfully shown in this
matrix model that the SO(4) symmetry is broken spontaneously down to SO(2).
However, the validity of the extrapolation relies on the assumption
that there is no phase transition in the region that cannot be studied
by the CLM. Also it is not straightforward to estimate
the systematic error associated with the extrapolation.
In order to address these important issues in the deformation technique,
we consider 
three types of deformation of the Dirac operator and compare the result
after extrapolations.

\begin{figure}[tb]
\centering
\includegraphics[width=6.7cm,bb=30mm 15mm 95mm 85mm,clip]{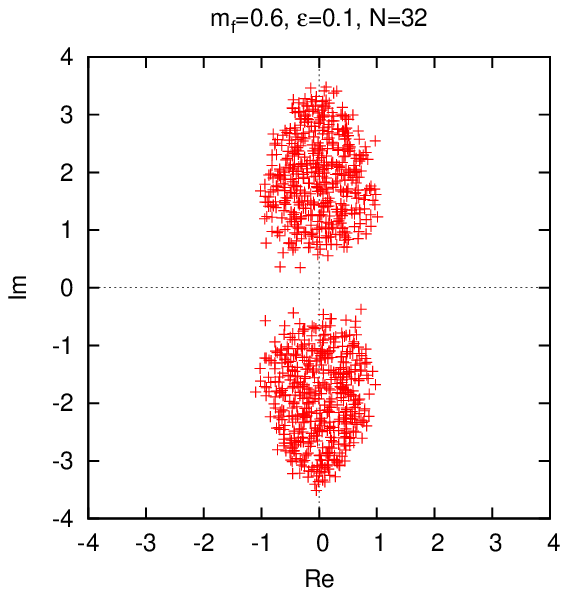}\includegraphics[width=6.7cm,bb=30mm 15mm 95mm 85mm,clip]{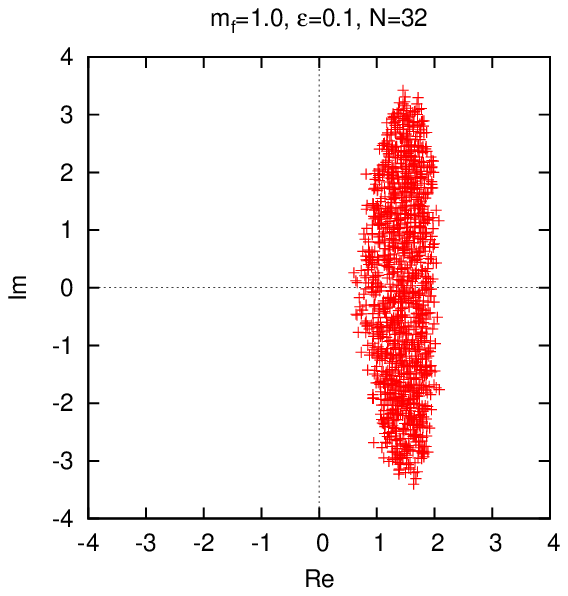}
\caption{(Left) The scatter plot of the eigenvalues of the deformed Dirac
operator (\ref{eq:deform_1A}) with $\varepsilon=0.1$, $m_{\mathrm{f}}=0.6$
and $N=32$. (Right) The scatter plot of the eigenvalues of the
deformed Dirac operator (\ref{eq:deform_1B}) with $\varepsilon=0.1$,
$m_{\mathrm{f}}=1.0$ and $N=32$.
In both plots, the configuration was obtained by simulating the corresponding
deformed model.
\label{fig:scattered-plot}}
\end{figure}

The first two types are the ones we have used in
our previous work \cite{Ito:2016efb},
which are defined by
\begin{equation}
D_{1A}' = 
X_{\mu}\otimes\Gamma_{\mu}+m_{\mathrm{f}}
\mathbf{1}_{N}\otimes\Gamma_{3} \ ,
\label{eq:deform_1A}
\end{equation}
\begin{equation}
D_{1B}' = X_{\mu}\otimes\Gamma_{\mu}+m_{\mathrm{f}}
\mathbf{1}_{N}\otimes\Gamma_{4} \ ,
\label{eq:deform_1B}
\end{equation}
where we added the second term to the original Dirac operator (\ref{eq:dirac_op})
and $m_{\mathrm{f}}$ is the deformation parameter. 
For sufficiently large $m_{\mathrm{f}}$, the second
term changes the eigenvalue distribution
of the Dirac operator in such a way that 
there are no near-zero eigenvalues. 
In figure \ref{fig:scattered-plot}, we plot the eigenvalue distribution
of the deformed Dirac operators (\ref{eq:deform_1A}) and (\ref{eq:deform_1B})
obtained by simulating the corresponding deformed models.
We find that the appearance of near-zero eigenvalues
is avoided even for $\varepsilon=0.1$. 
For a fixed $m_{\mathrm{f}}$, we can extrapolate
$\varepsilon$ to zero using the results obtained in the large-$N$ limit.
We can then extrapolate the deformation
parameter $m_{\mathrm{f}}$ to zero to get the results for the original model.

\begin{figure}[tb]
\centering
\includegraphics[width=6.7cm,bb=30mm 15mm 95mm 85mm,clip]{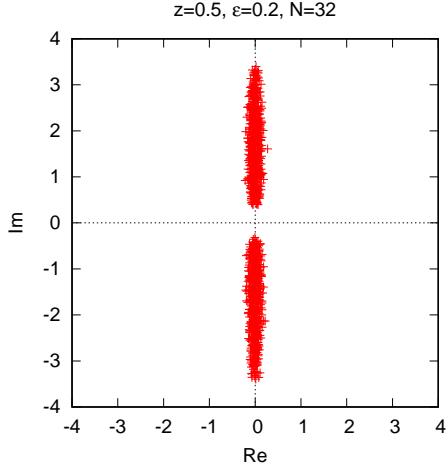}
\caption{The scatter plot of the eigenvalues of the deformed Dirac
operator (\ref{eq:deform_2}) with $\varepsilon=0.2$, $z=0.5$
and $N=32$. The configuration was obtained by simulating the corresponding
deformed model.
\label{fig:scattered-plot_deform2}}
\end{figure}

In this work we consider the third type of deformation, which is defined as
\begin{equation}
D_{2}'= \sum_{i=1}^3 X_{i}\otimes\Gamma_{i} + z X_{4}\otimes\Gamma_{4} \ ,
\label{eq:deform_2}
\end{equation}
where we have introduced a deformation parameter $z$.
The original model corresponds to $z=1$
and the SO(4) symmetry is broken explicitly to SO(3) for $z\neq 1$.
Note that this deformed Dirac operator
$D_{2}'$ becomes anti-Hermitian when $z$ is pure imaginary,
in which case $\det D_{2}'$ is real and there is no sign problem.
In this sense the deformation parameter $z$ may be viewed as
an analogue of the chemical potential in finite density QCD. 
The singular-drift problem is expected
not to occur when $|\mathrm{Re}\,z|$ is small.
In figure \ref{fig:scattered-plot_deform2}, we plot the eigenvalue distribution
of the deformed Dirac operator (\ref{eq:deform_2})
obtained by simulating the corresponding deformed model.
We find that the appearance of near-zero eigenvalues
is avoided even for $\varepsilon=0.2$. 
Note that for $z=0$, all the eigenvalues lie on 
the imaginary axis excluding the region near the origin.
In what follows we restrict ourselves to the case with a real $z$
for simplicity and see
whether the SO(3) symmetry of the deformed model
is spontaneously broken down to SO(2)
as we increase $z$ towards $z=1$.
We also try to make an extrapolation to $z=1$.

\section{Results\label{sec:Results}}

Before presenting our results, let us define the ratios 
\begin{equation}
\rho_{\mu}\left(\varepsilon,m_{\mathrm{f}}\right)
=\lim_{N\rightarrow\infty}
\frac{\left\langle \frac{1}{N}\mathrm{tr}X_{\mu}^{2}
\right\rangle _{\varepsilon,m_{\mathrm{f}}}}{\sum_{\nu=1}^{4}
\left\langle \frac{1}{N}\mathrm{tr}X_{\nu}^{2}\right\rangle 
_{\varepsilon,m_{\mathrm{f}}}} 
\label{eq:ratio}
\end{equation}
corresponding to the deformation (\ref{eq:deform_1A})
and similarly for the other deformations
(\ref{eq:deform_1B}) and (\ref{eq:deform_2}). 
This is motivated from the fact that the
mass term (\ref{eq:boson_action_bdeform}) tends to make all the expectation
values $\langle\lambda_{\mu}\rangle_{\varepsilon,m_{{\rm f}}}$ smaller
than the value to be obtained in the $\varepsilon\rightarrow0$ limit.
By taking the ratio (\ref{eq:ratio}), the finite $\varepsilon$ effects
are canceled by the denominator, and the extrapolation to $\varepsilon=0$
becomes easier.

\begin{figure}[tb]
\centering
\includegraphics[width=7.2cm,clip]{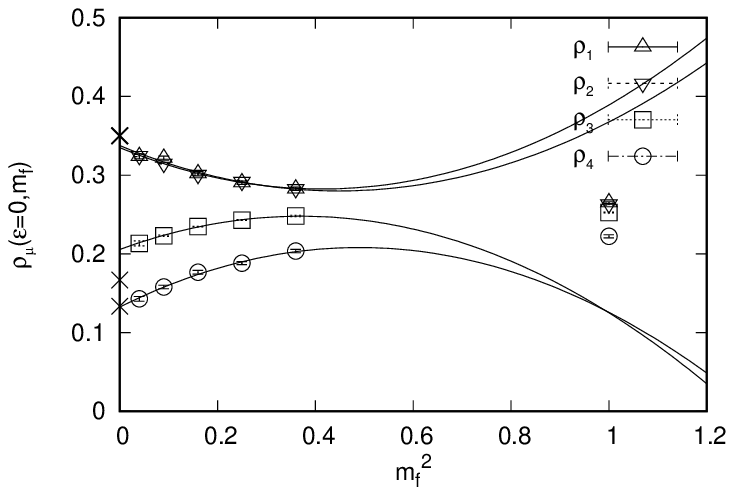}\includegraphics[width=7.2cm,clip]{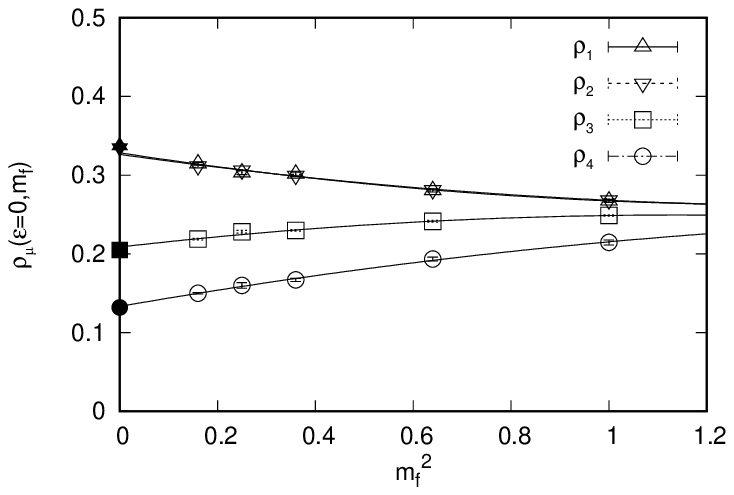}
\caption{(Left) The ratio $\rho_{\mu}\left(0,m_{\mathrm{f}}\right)$ obtained
after taking the large-$N$ limit and the $\varepsilon\rightarrow0$ limit
in the deformed model defined by (\ref{eq:deform_1A})
is plotted against $m_{\mathrm{f}}^{2}$. The lines represent fits
to the quadratic form $a+bm_{\mathrm{f}}^{2}+cm_{\mathrm{f}}^{4}$.
The crosses at $m_{\mathrm{f}}=0$ represent the prediction
from the GEM. 
(Right) The ratio $\rho_{\mu}\left(0,m_{\mathrm{f}}\right)$
obtained in the deformed model defined by (\ref{eq:deform_1B})
is plotted against $m_{\mathrm{f}}^{2}$.
The lines represent fits to the quadratic form 
$a+bm_{\mathrm{f}}^{2}+cm_{\mathrm{f}}^{4}$.
The filled symbols at $m_{\mathrm{f}}=0$ represent the results
obtained with the deformation (\ref{eq:deform_1A}).
\label{fig:ratio def_1A1B}}
\end{figure}

In figure \ref{fig:ratio def_1A1B} (Left), 
we plot the ratio $\rho_{\mu}\left(0,m_{\mathrm{f}}\right)$
after taking the large-$N$ limit and the $\varepsilon\rightarrow0$ limit
against $m_{\mathrm{f}}^{2}$ for the deformation (\ref{eq:deform_1A}).
The lines represent fits to the form $a+bm_{\mathrm{f}}^{2}+cm_{\mathrm{f}}^{4}$
using the data for $0.2\leq m_{\mathrm{f}}\leq0.6$. 
By extrapolating $m_{\mathrm{f}}$ to 0, we obtain
results for the original model. In figure
\ref{fig:ratio def_1A1B} (Left), we also plot the prediction from
the GEM with crosses at $m_{\mathrm{f}}=0$,
which reveal certain discrepancies from the results
obtained by the extrapolation with the deformation (\ref{eq:deform_1A}).

In figure \ref{fig:ratio def_1A1B} (Right), 
we present a similar plot for the deformation (\ref{eq:deform_1B}).
The lines represent fits to the form $a+bm_{\mathrm{f}}^{2}+cm_{\mathrm{f}}^{4}$
using the data for $0.4\leq m_{\mathrm{f}}\leq1.0$. 
In the same figure,
we also plot the results obtained with the deformation (\ref{eq:deform_1A}) 
with filled symbols at $m_{\mathrm{f}}=0$,
which are found to be in good agreement
with the results obtained with the deformation (\ref{eq:deform_1B}).
This implies that the discrepancies between
our results and the prediction from the GEM 
are due to the approximation used in GEM.

\begin{figure}
\centering
\includegraphics[width=9.0cm,clip]{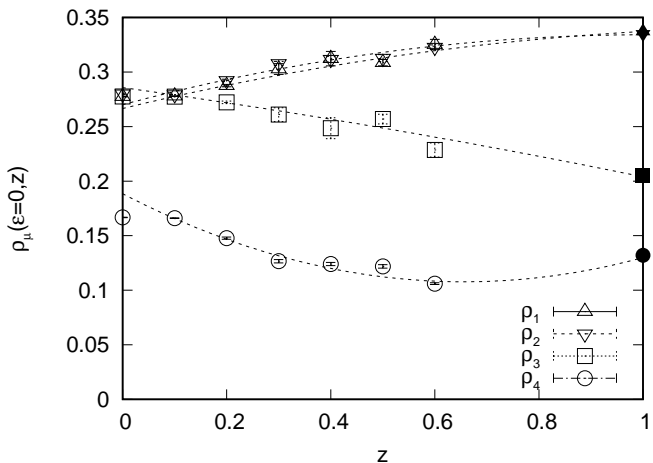}
\caption{The ratio $\rho_{\mu}\left(0,z\right)$ obtained after taking 
the large-$N$ limit and the $\varepsilon\rightarrow0$ limit 
in the deformed model defined by (\ref{eq:deform_2}) 
is plotted against $z$. The filled symbols at $z=1$ represent the results
obtained with the deformation (\ref{eq:deform_1A}). 
The dashed lines represent fits to the quadratic
form $a+bz+cz^{2}$ using the data for $0.2\leq z\leq0.6$ 
and the data points at $z=1$, which is obtained 
with the deformation (\ref{eq:deform_1A}).
\label{fig:ratio-def_2}}
\end{figure}

In figure \ref{fig:ratio-def_2}, we show our preliminary results
for the deformation (\ref{eq:deform_2}).
We find that the SO(3) symmetry of the deformed model
is spontaneously broken down to SO(2) for $z \gtrsim 0.2$.
We have also attempted to make an extrapolation to $z=1$
with the quadratic form $a+bz+cz^{2}$ using
the data for $0.2\leq z\leq0.6$, but
the uncertainties of our present data
prevent us from making a reliable extrapolation at this stage.
Here we would like to 
content ourselves with just checking the consistency with
the deformation (\ref{eq:deform_1A}).
For that purpose,
we fit the data including the
data points at $z=1$ obtained with the deformation (\ref{eq:deform_1A}).
The result of the fits looks reasonable 
as is shown in figure \ref{fig:ratio-def_2} with dashed lines.
This suggests that it is possible to improve the data so that we can make 
a reliable extrapolation without using the data points at $z=1$.

\section{Summary and Discussions\label{sec:Discussion}}

In this article we discussed the deformation technique, which
enables us to avoid the singular-drift problem that occurs in the CLM.
We investigated the matrix model with a complex
fermion determinant, in which the SO(4) symmetry is
expected to be broken spontaneously down to SO(2) due to the effect of 
the phase of the fermion determinant. In order to avoid the singular-drift
problem, we deformed the Dirac operator in three different
ways so that the appearance of near-zero eigenvalues is suppressed.
Our results suggest that
the final results are independent of how we deform the Dirac operator.
The deviations from the prediction by the GEM are therefore attributed to
the approximation used in the GEM.
The systematic errors associated with the extrapolation are considered
to be smaller than these deviations.
We hope that the insight gained in this work will be useful in 
applying the same technique to the complex Langevin analysis 
of finite density QCD at high density and low temperature \cite{shimasaki}.

In fact, the matrix model investigated in this work may be viewed
as a toy model for the matrix model conjectured to 
be a non-perturbative formulation of the
type IIB superstring theory in ten dimensions \cite{Ishibashi:1996xs}. 
In that model, the SO(10) symmetry is expected to be broken down to SO(4)
in order to account for our four-dimensional space-time.
On the other hand, the GEM 
predicts that the SO(10) symmetry is broken down to SO(3) 
instead of SO(4) \cite{Nishimura:2011xy}.
It would be interesting to address this issue
extending the present work.

\section*{Acknowledgements }

The authors would like to thank K.N.~Anagnostopoulos, T.~Azuma,
K.~Nagata, S.K.~Papadoudis and S.~Shimasaki for valuable discussions.
Y.~I.\ is supported by JICFuS. J.~N.\ is supported in part by
Grant-in-Aid for Scientific Research (No.\ 16H03988)
from Japan Society for the Promotion of Science.

\end{document}